\documentclass{article}
\usepackage{graphicx}
\usepackage{amsmath}
\usepackage{amsfonts}
\usepackage{amssymb,bbm}
\usepackage{url}

\usepackage{enumitem}

\usepackage{xcolor}

\newcommand{\ZZ}{{\mathbbm{Z}}}

\newcommand{\N}{\mathbb{N}}

\newcommand{\bb}{{\mathbf b}}

\newcommand{\tabD}[8]{\begin{tabular}{c | c c c c c c c c}
$xyz$ & 000 & 001 & 010 & 011 & 100 & 101 & 110 & 111\\
$ f(x,y,z) $ &  #1 & #2 & #3 & #4 & #5 & #6 & #7 & #8\\
\end{tabular}
}
\newcommand{\tabP}[8]{\begin{tabular}{c | c c c c c c c c}
		$ xyz$ & 000 & 001 & 010 & 011 & 100 & 101 & 110 & 111\\
		$ w(1 | x,y,z) $ &  $#1$ & $#2$ & $#3$ & $#4$ & $#5$ & $#6$ & $#7$ & $#8$\\
	\end{tabular}
}

\begin{document}

\title{Analysis of the Asymptotic Density of Endogamous Diploid Cellular Automata}

\author{
Henryk Fukś
\footnote{\texttt{hfuks@brocku.ca} ; Department of Mathematics and Statistics, Brock University,
St. Catharines, ON L2S 3A1, Canada} and Nazim Fatès
\footnote{\texttt{nazim.fates@inria.fr};
Université de Lorraine, CNRS, Inria, LORIA, Nancy, F-54000,  France}
}

\date{June 2026}

\maketitle
\abstract{
We investigate the asymptotic behaviour of diploid Elementary Cellular Automata (ECA), that is, the stochastic mixtures between two ECAs.
In this model, each cell independently applies one rule with probability $\lambda$ and the other rule with probability $ 1 - \lambda$. 
Focusing on the endogamous diploids where the two ECAs are related by the reflection or conjugation symmetry,
we analyse how the density varies as function of $\lambda$, the ``degree of mixing'' of these two rules.
We propose a classification into six distinct classes depending on the profile of the density vs. $\lambda$ curve. 
We take various examples for each class and we analyse to which extent the local structure approximation succeeds to predict the asymptotic density.

Our results show that for rules in which the asymptotic density depends linearly on $\lambda$, the local structure approximation reproduces the exact dependence of the  density on $\lambda$. For rules with differentiable but nonlinear dependence, the approximation either becomes exact at a finite order or converges rapidly to the exact solution as the order increases.
For rules with first-order phase transitions, finite-order approximations  progressively approach a sharp transition profile with increasing order.
In contrast, for rules exhibiting a second-order phase transition, even high-order approximations fail to capture the qualitative features of the transition. (no bifurcation is observed). 
Finally, we give examples of two diploid rules for which we succeed to compute the density at each time step.
}

\section{Introduction}

Probabilistic cellular automata (PCA) remain a largely unexplored field. They have been used to model a wide range of natural phenomena and are also of independent interest as simple models of dynamical systems, in which complex behaviour emerges from simple local interactions. Moreover, it has been shown that there are different computational problems for which the randomness inherent to these systems can be an aid to attain a solution (see e.g. Ref.~\cite{RR18}).

PCA are frequently used for modelling various biological phenomena (see e.g. Ref.~\cite{DeuDor05}). 
In her work, D. Makowiec developed several PCA models to model the dynamics of excitation waves in the heart~\cite{Mak10,MWS19}.
She also studied a two-dimensional PCA known as the probabilistic
Toom's rule~\cite{Makowiec1997PRE,Makowiec1998APPB}. The main subject of her investigations  for this rule were its  stationary states, as well as decay of correlations, boundary configuration dependence, and large-scale fluctuations.

In this article, we focus on the space of  {\em elementary} PCA, which is isomorphic to $[0,1]^8$, as each PCA rule can be described by eight probabilities. Although quite simple in appearance, the space of PCA rules is far from being fully explored and there are many simple questions which are still open (see e.g., Ref.~\cite{CMP26}).
As a starting point, one can consider {\em diploid CA}, where two deterministic Elementary Cellular Automata (ECA) are mixed stochastically~\cite{Fdip17}. However, this class of rules is still very large as it contains 8808 rules. This space is too large to be studied exhaustively and, recently, Roy \emph{et al.} considered the case where the two ECAs are obtained by an arbitrary ECA and one of its symmetrical rules~\cite{RMPA24}. Although their approach was mainly relying on empirical observations, they showed that this restricted set of rules was still rich enough to display many interesting phenomena. 
In a recent work, we made a first step for analysing mathematically this subset of 168 PCA rules, called {\em endogamous rules}, and we proposed a tentative classification of their different behaviours according to the average convergence time to a fixed point~\cite{Fendo25}.

Among the few works devoted to this subset of PCA rules, we mention the study of Bo{\l}t et al., who tackled the {\em identification problem}, that is, the task to reconstruct the original rule from partial observations of their space-time diagrams~\cite{BBWBdB19}.
Cirillo \emph{et al.} made an analysis of the diploids which are mixed with the null rule~\cite{CNS21}. 
They studied how local structure theory, or block approximations, can predict the asymptotic density of these rules~\cite{CLS24}.
In a recent paper, they examined in great detail a specific endogamous rule, namely the couple 60-102, to study to which extent the block approximation fails to predict the essential features of the phase transition that appears when one changes the weight of each rule~\cite{CVSS26}. They also rigorously prove that for the particular value where the two rules have an equal probability to be applied ($\lambda=1/2$), the attractive steady-state consists of the all-zero stable configuration.

In this paper, we will restrict 
our investigations to endogamous rules but note that the techniques readily apply to all the diploid rules.
We want to examine to which extent the asymptotic density can be predicted analytically, where by the asymptotic density we mean the fraction of sites in state 1 in the steady state. 
 We believe that the accuracy of such a prediction somehow measures how complex the behaviour of a rule is. We will propose a qualitative classification of the rules based on numerical simulations and then try to relate the different classes with their ability to 'resist' an effective prediction of their steady-state by the simple means of mean-field approximation or by the more elaborate technique of local structure theory (also called block approximation).

The paper is organized as follows. In the next section we provide the necessary background and definitions of
Elementary Cellular Automata, diploid, and endogamous rules.
Section 3 describes results of numerical experiments which 
are presented by grouping endogamous rules into six classes
based on the behaviour of the asymptotic density as a function of $\lambda$.  
For each class, we select a representative example and investigate how well the mean-field approximation
or its generalization can predict the shape of the
asymptotic density versus $\lambda$ curve. In Section~4
we present further two examples of rules which
are exactly solvable in the sense that the asymptotic density
can be computed analytically. Our findings are then summarized in the Conclusion section. 
\section{Basic definitions}

We will consider one-dimensional binary cellular automata for 
which
the set of states that a cell can assume is $ \{ 0, 1 \}$. We will denote by $s_i(t)$  the state of the lattice cell $i \in \ZZ $ at time $t \in \N$. In the diploid CA which we consider in this paper, 
$s_i(t)$ is a random variable, but we  first 
need to define a purely deterministic CA.
A \emph{deterministic Elementary Cellular Automaton} is a dynamical system 
governed by a local function $f: \{0,1\}^3 \to \{0,1\}$ such that 
$  \forall t \in \mathbb{N}, \forall i \in \mathbb{Z},$
$$
 s_i(t+1)=f\big(s_{i-1}(t), s_{i}(t), s_{i+1}(t)\big).
$$
The function $f$ is  called the \emph{local rule}, or for short, {\em the rule} of the CA.

Elementary rules are usually identified
by their Wolfram number $W_f $, defined as 
$ W_f= f(0,0,0) \cdot 2^0 +  f(0,0,1) \cdot 2^1 + \dots + f(1,1,1) \cdot 2^7$.
The values of the local function $f$ can thus be constructed by converting $W_f$ to a binary representation. Consecutive bits, read from 
right to left, correspond to values of $f(x_1,x_2,x_3)$ arranged in lexicographical order, $f(0,0,0)$,  $f(0,0,1), \ldots,$
$f(1,1,1)$. For example, for the elementary rule with 
Wolfram number 140 we have
$$
W_f= 140=(10001100)_2
$$
hence the following transition table:
\vspace{4pt}

\tabD 0 0 1 1 0 0 0 1.

\smallskip
There are 256 ECA and some rules are related by a symmetry transformation.
One of such symmetries is the Boolean conjugation $C$, defined as 
$$
   Cf(x_1,x_2,x_3)=1-f(1-x_1,1-x_2,1-x_3)
  \text{\,\,\,for any $f:\{0,1\}^3 \to \{0,1\}$}.
$$  
Another symmetry is the spatial reflection $R$
 which interchanges the
roles of the left and the right. The action of $R$ on local functions
is defined by
$$  Rf(x_1,x_2, x_3)=f(x_3,x_{2}, x_1)
  \text{\,\,\,for any $f:\{0,1\}^3 \to \{0,1\}$.} \nonumber
$$
We will also  denote by $I$ the identity operator which does not change
the local function, $If=f$. One can easily show that 
 $\{I,R,C,RC\}$ form an abelian group.
This means that given the local function $f$, there
are only four varieties of local functions which one can obtain by repeated
application of $R$, $C$ or any combination of thereof, namely
$f$, $Rf$, $CF$ and $RCf$. Depending on $f$, some of these may be identical.
The set of  four rules $\{If, Cf, Rf, RCf\}$, after removal of duplicates,
is called the \emph{equivalence class} of CA rules with respect to
the aforementioned group of symmetries. It turns out that there are
88 of such equivalence classes, shown in Table~\ref{minimalrules}.
The rule with the smallest Wolfram number is usually used as the 
representative of the class -- minimal representatives are shown in 
bold in Table~\ref{minimalrules}.
 \begin{table}
    \setlength{\tabcolsep}{6pt} 
    \renewcommand{\arraystretch}{1.2}
     \caption[Table of the 88 equivalence classes of  Elementary Cellular
    Automata.]{Table of the 88 equivalence classes of  Elementary
    	Cellular Automata. Rules with the minimal Wolfram number
    	in each class are shown in bold.}
    	\label{minimalrules}
    \begin{center}
      {\footnotesize
        \begin{tabular}{|l|l|l|l|}
          $\{\mathbf{0},255          \}$ &
          $\{\mathbf{26},82,167,181  \}$  &
          $\{\mathbf{56},98,185,227  \}$  &   $\{\mathbf{132},222
          \}$   \\
          $\{\mathbf{1},127          \}$ &   $\{\mathbf{27},39,53,83
          \}$  &   $\{\mathbf{57},99          \}$  &
          $\{\mathbf{134},148,158,214  \}$   \\
          $\{\mathbf{2},16,191,247   \}$ &
          $\{\mathbf{28},70,157,199  \}$  &
          $\{\mathbf{58},114,163,177 \}$  &
          $\{\mathbf{136},192,238,252  \}$   \\
          $\{\mathbf{3},17,63,119    \}$ &   $\{\mathbf{29},71
          \}$  &   $\{\mathbf{60},102,153,195 \}$  &
          $\{\mathbf{138},174,208,244  \}$   \\
          $\{\mathbf{4},223          \}$ &
          $\{\mathbf{30},86,135,149  \}$  &
          $\{\mathbf{62},118,131,145 \}$  &
          $\{\mathbf{140},196,206,220  \}$   \\
          $\{\mathbf{5},95           \}$ &   $\{\mathbf{32},251
          \}$  &   $\{\mathbf{72},237         \}$  &
          $\{\mathbf{142},212          \}$   \\
          $\{\mathbf{6},20,159,215   \}$ &   $\{\mathbf{33},123
          \}$  &   $\{\mathbf{73},109         \}$  &
          $\{\mathbf{146},182          \}$   \\
          $\{\mathbf{7},21,31,87     \}$ &
          $\{\mathbf{34},48,187,243  \}$  &
          $\{\mathbf{74},88,173,229  \}$  &   $\{\mathbf{150}
          \}$   \\
          $\{\mathbf{8},64,239,253   \}$ &   $\{\mathbf{35},49,59,115
          \}$  &   $\{\mathbf{76},205         \}$  &
          $\{\mathbf{152},188,194,230  \}$   \\
          $\{\mathbf{9},65,111,125   \}$ &   $\{\mathbf{36},219
          \}$  &   $\{\mathbf{77}             \}$  &
          $\{\mathbf{154},166,180,210  \}$   \\
          $\{\mathbf{10},80,175,245  \}$ &   $\{\mathbf{37},91
          \}$  &   $\{\mathbf{78},92,141,197  \}$  &
          $\{\mathbf{156},198          \}$   \\
          $\{\mathbf{11},47,81,117   \}$ &
          $\{\mathbf{38},52,155,211  \}$  &   $\{\mathbf{90},165
          \}$  &   $\{\mathbf{160},250          \}$   \\
          $\{\mathbf{12},68,207,221  \}$ &
          $\{\mathbf{40},96,235,249  \}$  &   $\{\mathbf{94},133
          \}$  &   $\{\mathbf{162},176,186,242  \}$   \\
          $\{\mathbf{13},69,79,93    \}$ &
          $\{\mathbf{41},97,107,121  \}$  &   $\{\mathbf{104},233
          \}$  &   $\{\mathbf{164},218          \}$   \\
          $\{\mathbf{14},84,143,213  \}$ &
          $\{\mathbf{42},112,171,241 \}$  &   $\{\mathbf{105}
          \}$  &   $\{\mathbf{168},224,234,248  \}$   \\
          $\{\mathbf{15},85          \}$ &   $\{\mathbf{43},113
          \}$  &   $\{\mathbf{106},120,169,225\}$  &
          $\{\mathbf{170},240          \}$   \\
          $\{\mathbf{18},183         \}$ &
          $\{\mathbf{44},100,203,217 \}$  &   $\{\mathbf{108},201
          \}$  &   $\{\mathbf{172},202,216,228  \}$   \\
          $\{\mathbf{19},55          \}$ &   $\{\mathbf{45},75,89,101
          \}$  &   $\{\mathbf{110},124,137,193\}$  &
          $\{\mathbf{178}              \}$   \\
          $\{\mathbf{22},151         \}$ &
          $\{\mathbf{46},116,139,209 \}$  &   $\{\mathbf{122},161
          \}$  &   $\{\mathbf{184},226          \}$   \\
          $\{\mathbf{23}             \}$ &   $\{\mathbf{50},179
          \}$  &   $\{\mathbf{126},129        \}$  &
          $\{\mathbf{200},236          \}$   \\
          $\{\mathbf{24},66,189,231  \}$ &   $\{\mathbf{51}
          \}$  &   $\{\mathbf{128},254        \}$  &
          $\{\mathbf{204}              \}$   \\
          $\{\mathbf{25},61,67,103   \}$ &   $\{\mathbf{54},147
          \}$  &   $\{\mathbf{130},144,190,246\}$  &
          $\{\mathbf{232}              \}$   \\
      \end{tabular} }
    \end{center}
   
  \end{table}

In a  \emph{probabilistic cellular automaton} (PCA), states of lattice cells 
$s_i(t)$ are random variables
taking values  in $\{0,1\}$. The value of $s_i(t+1)$ depends 
on probabilities which are calculated according to the states of local neighbours. This is usually expressed using 
transition probabilities $w(a|\bb)$, $a\in \{0,1\}$, $\bb \in \{0,1\}^3$, as follows,
\begin{equation}\label{defw}
Pr(s_i(t+1)=a)=w(a|s_{i-1}(t),s_{i}(t), s_{i+1}(t)).
\end{equation}

Note that $w(1|\bb)+w(0|\bb)=1$ for all $\bb
\in \{0,1\}^3$. For this reason, the nearest-neighbour PCA rule
can be defined with the eight transition probabilities $w(1|x_1 x_2x_3)$ for all $x_1 x_2 x_3 \in \{0,1\}$. 
The remaining eight probabilities, $w(0|x_1 x_2 x_3)$ for all $x_1,x_2,x_3 \in \{0,1\}$, can be obtained by $w(0|x_1 x_2 x_3)=1-w(1|x_1x_2x_3)$.

\smallskip
We now define \emph{diploid elementary cellular automata}. 
Let $f$ and $g$ be local functions of two ECA with Wolfram numbers, respectively, $W_f$ and $W_g$ and
let $\lambda \in [0,1]$ be a parameter. 
We define a diploid rule, to be referred to as $W_f-W_g$,
to be the PCA rule with the transition probabilities given by 
\begin{equation} \label{tranprobdip}
 w(1|x_1 x_2 x_3)=(1-\lambda) f(x_1,x_2,x_3)+ \lambda g(x_1,x_2,x_3).
\end{equation}
Note that when $\lambda=0$ or $\lambda=1$, the diploid rule is equivalent to the deterministic rule with the local function $f$ or $g$, respectively.

As we already mentioned, elementary cellular automata can be divided into 88 equivalence classes
with  respect to the symmetry group $\{ I,C,R,CR \}$.
If $f$ and $g$ belong to the same equivalence class, we call the
$W_f-W_g$ rule an \emph{endogamous rule}. 
It is known 
that there are 168 endogamous rules which are truly distinct~\cite{Fendo25},
and these rules will be the subject of our study.
These are constructed by taking as $f$ a minimal rule
belonging to a given equivalence class,
and then as $g$ any of the remaining rules of the same class. Since there are 256 elementary CA, the number of
endogamous rules is $256-88=168$.

We will assume that the stochastic process with transition probabilities defined in eq.~(\ref{defw}) is iterated from an initial condition that is shift-invariant. By the properties of a PCA, it thus remains in distribution
shift-invariant over time. 
At time $ t$, the probability of occurrence of a symbol $b \in  \{0,1\} $ at cell $ i $ is thus independent of $ i $ and can be denoted by $ P_t(b)$, 
so that
\begin{equation}
P_t(b)=Pr\Big(s_i(t)=b \Big).
\end{equation}
We will be mainly interested in the expected value of cell $i$ at time 
$t$, which will be called the $\emph{density}$ and denoted by $d_t$, with
$$
d_t = \langle s_i(t) \rangle = 1 \cdot P_t(1)+0 \cdot P_t(0)=
P_t(1).
$$
The density $d_t$, therefore, is the same as the probability of occurrence of 1. 

Suppose that we start with a symmetric Bernoulli distribution, so that the lattice cells are randomly and independently set to  state 0 or 1 with probabilities $1/2$, meaning that  $d_0=0.5$. 
The ``steady state'' density will be denoted by
$$
d_\infty = \lim_{t \to \infty} d_t.
$$

Note that $d_t$ and $d_\infty$ as defined above are 
\emph{theoretical} expected values of $s_i(t)$, for infinite lattice and assuming initial symmetric 
Bernoulli measure. In what follows, we will also use 
\emph{empirical density} obtained in numerical experiments, where, for a finite configuration of length $L$ with periodic boundary conditions,
$$\tilde{d}_t=\frac{1}{L} \sum_{i=0}^{L-1} s_i(t).$$
 By the law of large numbers, for very large lattices  one can expect that the empirical densities  should be close to
theoretical ones, but obviously the correspondence is only approximate for a finite $ t $ and, as $ t$ grows, finite-size effects might become significant.
Nevertheless, for the sake of simplicity, we will use the same symbol $d_t$ for both theoretical and empirical densities as it will always be clear from the context which one is used.

The central question we want to ask is as follows: for endogamous rules, how does the steady state density depend on the parameter $\lambda$?
\section{Numerical experiments and  mean-field approximation}

In order to answer the question above, we first 
constructed $d_\infty$ vs. $\lambda$ graphs by numerical simulations,
by iterating all the 168 endogamous rules starting from a random initial configuration of length $L$ (with periodic boundary conditions) for $T$ time steps and measuring the final density. In order to reduce the noise and improve the quality of the graphs, we repeated each simulation $N$ times to obtain the average density. 
The parameter $\lambda$  was varied from 0 to 1 with increments of $0.02$.
The values used in our simulations were
$L=500$, $T=10^4$ and $N=4000$. We found that further increase of $T$ does not
cause any perceptible changes into the graphs, thus $T=10^4$ is sufficient
for the purpose of our investigations.

\begin{figure}
 \begin{center}
\includegraphics[width=5.7cm]{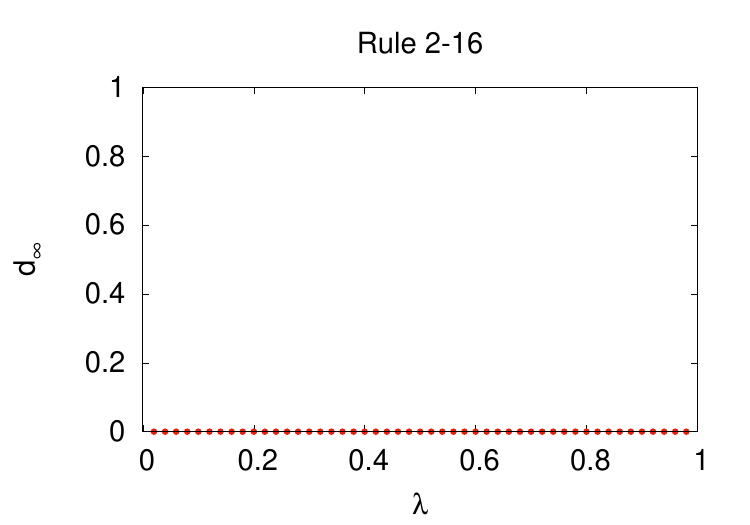}
\includegraphics[width=5.7cm]{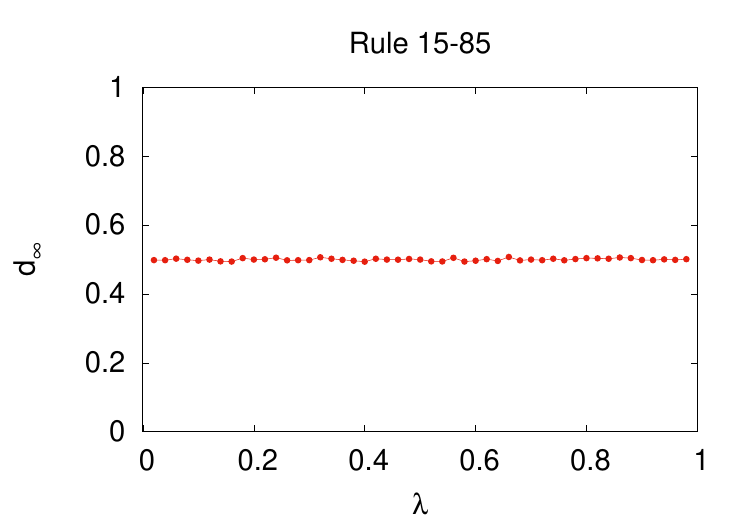}\\
\includegraphics[width=5.7cm]{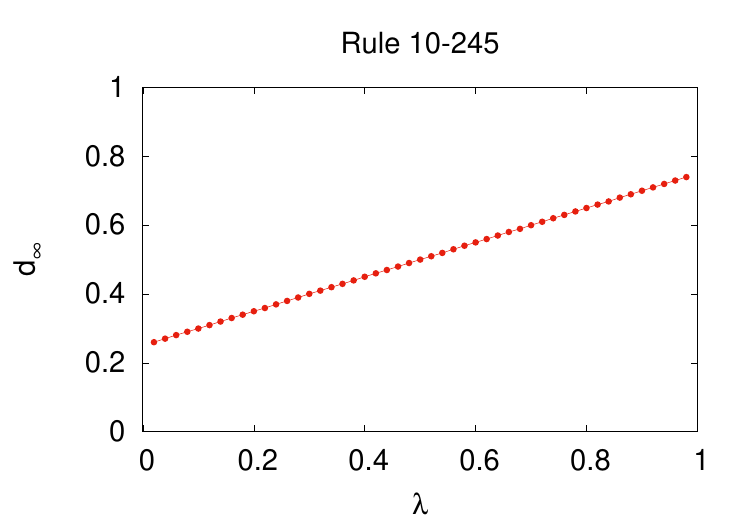}
\includegraphics[width=5.7cm]{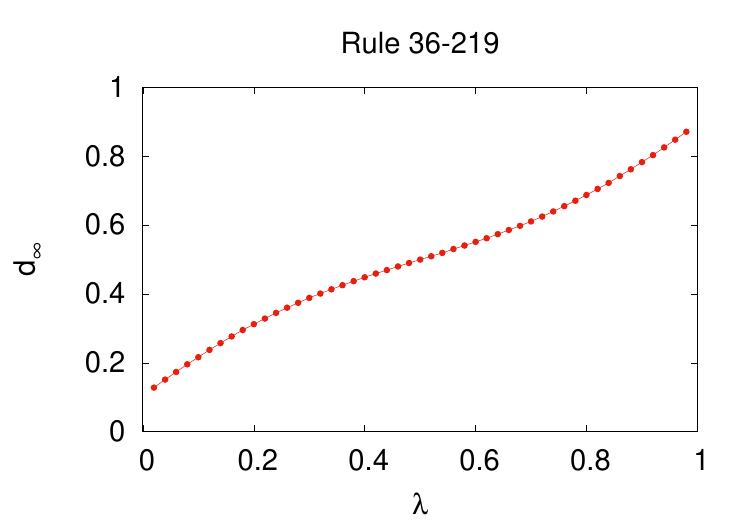}\\
\includegraphics[width=5.7cm]{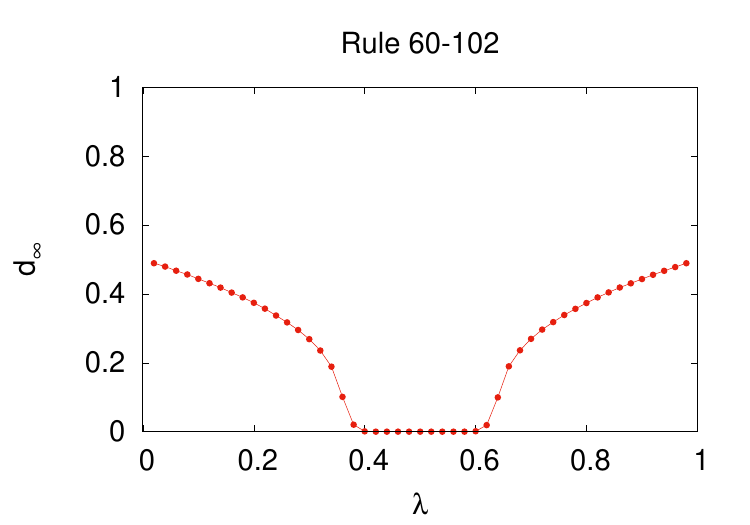}
\includegraphics[width=5.7cm]{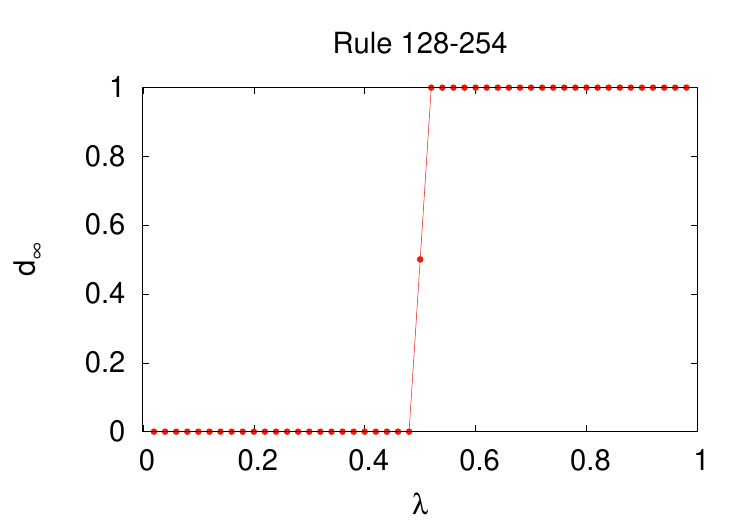}
 \end{center}
\caption{Examples of graphs of the steady-state density vs. $\lambda$
for six classes of endogamous  diploid rules.}\label{Fig-densexample}
\end{figure}

To verify that periodic boundary conditions do not introduce any
artifacts to the graphs, we performed simulations using larger lattice of size $L_\mathrm{big}=L+2T$ and then computing the density using only
central $L$ cells. Again, the graphs obtained this way do not show any noticeable difference from graphs shown in Figure~\ref{Fig-densexample}.

The graphs of $d_\infty$ versus $\lambda$ fall into six disjoint classes, shown in Table~\ref{table-classes}.
The labels of these classes are abbreviations of the respective property, ``zero density'' (ZD), ``constant density'', 
``linear density'' (LD), ``nonlinear density'' (NLD),
``phase transition of the second order'' (PT2)
and ''phase transition of the first order'' (PT1). 
We will explain further down why the term ``phase transition'' is used for classes PT1 and PT2.

Before we continue, two remarks must be made about the classification. First of all, this classification is
a vehicle to organize the graphs of of $d_\infty$ versus $\lambda$, but note that any set of functions $[0,1]\to [0,1]$
could be classified this way. Nevertheless, membership 
of endogamous rules in a given class may be rather hard to decide (or maybe even algorithmically undecidable in general). For example,
how do we know that a given graph is indeed exactly linear or just very close to linear? For this reason, we do not provide
a complete table of 168 rules showing to which class
each of them belongs. We only show examples for which,
based on extensive numerical simulations, we are reasonably confident about their classification.  These representatives examples are listed in the last column of Table~\ref{table-classes}.

The second remark we need to make is about the ZD
class. One could legitimately wonder why 
only $d_\infty=0$ rules are included but not 
rules for which $d_\infty=1$.
The reason for this is that
among 168 endogamous rules there is no case of $d_\infty=1$.
Recall that in every $W_f-W_g$ rule, $f$ is always minimal,
and by coincidence this requirement eliminates all the
 $d_\infty=1$ cases. 
\begin{table}
	\caption{Six classes of diploid rules.
		In the middle column, the given property is assumed to hold for all $\lambda\in (0,1)$}\label{table-classes}
\begin{tabular}{c|l|l}
Name & Definition & Example(s) \\ \hline \hline
ZD     &  $d_\infty=0$;                                                                              &  2-16 \\ \hline
CD     &  $d_\infty=\text{const} \neq 0$;                                                         & 15-85\\ \hline
LD     &  $d_\infty$  is a linear function of $\lambda$ but not constant;                         &  10-245\\ \hline
NLD    &  $d_\infty$  is a differentiable function of $\lambda$ but not linear;                   &    36-219\\ \hline
PT2    &  $d_\infty$ is a continuous function of $\lambda$ but not everywhere differentiable;     &   60-102 and 46-116 \\ \hline
PT1    &  $d_\infty$ is discontinuous at some point.                                              &  128-254 \\ \hline
\end{tabular}                                                                                                  

\end{table}

Graphs of rules selected as examples are shown in  Figure~\ref{Fig-densexample}.
Note that these graphs exhibit some obvious symmetries. 
If in the pair $W_f-Wg$ the rule $g$ is a spatial reflection of $f$, then the graph is symmetric with respect to the vertical line $\lambda=1/2$.
If $g$ is a conjugated or conjugated-reflected version of $f$, then the graph is 
symmetric with respect to the point $(1/2, 1/2)$.

In what follows, we will discuss these representative examples in detail.
We will compare the graphs obtained in numerical simulations with
approximate values of $d_\infty$ obtained from the
\emph{mean-field approximation}. If the mean-field approximation does not
produce results close enough to numerical results, we will resort to 
its generalization known as the
\emph{local structure approximation}. 
Exact details of constructions of the mean-field approximation 
and the local structure approximation for binary nearest-neighbour PCA
have been described in a previous work~\cite{hfpaper49}, we will thus skip these details here, giving only results. An expository
presentation of the local structure approximation can also be found in ch. 11 of \cite{hfbook2023}.

In Sec.~\ref{solvable}, we will discuss two additional examples of endogamous rules, namely 136-192 and 140-196, for which an exact expression for $d_t$ can be obtained.
These rules respectively belong to class ZD and NLD in the classification mentioned above.

\subsection{Rules 2-16 (class ZD) and 15-85 (class CD)}

Rules 2 and 16 are the reflected version of each other.
The diploid rule 2-16 is defined by the transition probabilities:

\vspace{4pt}
\tabP  0 {1 - \lambda} 0 0 \lambda 0 0 0 .

\smallskip
The mean-field approximation, 
as explained in \cite{hfpaper49}, is given by
$$
P_{t+1}(a)=
\sum_{b_1, b_2, b_3 \in \{0,1\}}w (a|b_1  b_2  b_3) P_t(b_1)
P_t(b_2) P_t(b_3).
$$
Taking $d_t=P_t(1)$ we obtain
\begin{equation}\label{mfeqgen}
d_{t+1}=
\sum_{b_1,b_2,b_3 \in \{0,1\}}w (1|b_1  b_2  b_3) P_t(b_1)
P_t(b_2) P_t(b_3),
\end{equation}
where we need to substitute $P_t(1)=d_t$ and 
$P_t(0)=1-d_t$.
For the diploid rule 2-16 this yields
\begin{equation}\label{mf2-16}
d_{t+1}=d_t^3-2d_t^2+d_t.
\end{equation}
Note that the equation does not involve $\lambda$. 
Indeed, rules 2 and 16 are related by the reflection transformation $R$.
More generally, if $g=Rf$, then $g(x_1,x_2,x_3)=f(x_3,x_2,x_1)$, and the transition probability defined in eq. (\ref{tranprobdip})  becomes
$$
w(1| b_1b_2b_3)=(1-\lambda)f(b_1,b_2,b_3) + \lambda
f(b_3,b_2,b_1).
$$
The mean-field equation (\ref{mfeqgen}) thus becomes
\begin{multline*}
d_{t+1}=
\sum_{b_1,b_2,b_3 \in \{0,1\}}w (1|b_1  b_2  b_3) P_t(b_1)
P_t(b_2) P_t(b_3)\\
=(1-\lambda) \sum_{b_1,b_2,b_3 \in \{0,1\}} f(b_1,b_2,b_3)
 P_t(b_1)
P_t(b_2) P_t(b_3)\\
+\lambda \sum_{b_1,b_2,b_3 \in \{0,1\}} f(b_1,b_2,b_3)
 P_t(b_1)
P_t(b_2) P_t(b_3),
\end{multline*}
hence
$$
d_{t+1}=\sum_{b_1,b_2,b_3 \in \{0,1\}} f(b_1,b_2,b_3)
 P_t(b_1)
P_t(b_2) P_t(b_3),
$$
where, again, we need to substitute $P_t(1)=d_t$ and 
$P_t(0)=1-d_t$.
The above equation is indeed independent of $\lambda$, and we will see this each time we consider an endogamous rule
constructed from $f$ and $Rf$.

\begin{figure}
		\begin{tabular}{ c c }
			rule 2-16 (ZD) & rule 15-85 (CD)\\
		\includegraphics[width=0.47\linewidth]{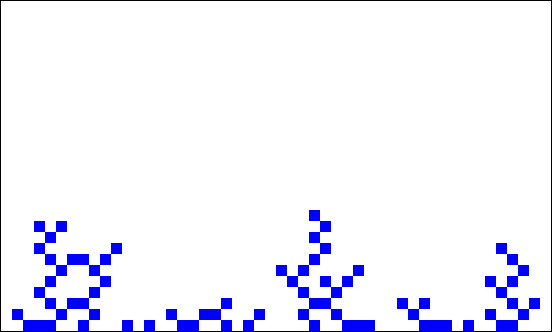}&
		\includegraphics[width=0.47\linewidth]{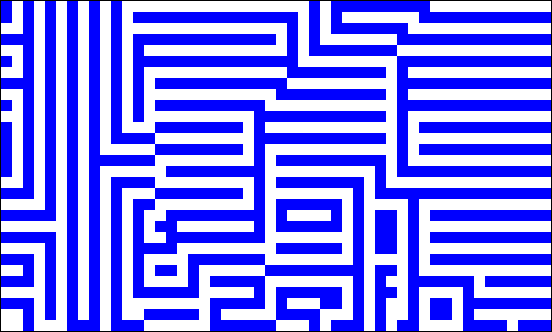}
	\end{tabular}
	\caption{Space-time diagrams for the rule 2-16 (left) and 15-85 (right) for $\lambda = 1/2$. Time goes from bottom to top. Blue/black and white cells represent cells in state 0 and 1, respectively. This convention is kept in the other figures.}
	\label{fig::stdZDCD}
\end{figure}

Coming back to eq. (\ref{mf2-16}): it has two fixed points, $d=0$ and $d=2$, but only  $d=0$ is stable, and this indeed is the steady state observed in Monte Carlo simulations. The observation of space-time diagrams confirm this tendency and reveals a quick convergence to the all-zero configuration (see Fig.~\ref{fig::stdZDCD}).

In the diploid 15-85, rules  15 and 85 are the reflected version of each other and are both invariant by conjugation. The transition probabilities
are
\vspace{4pt}
\tabP 1 {1 - \lambda} 1 {1 - \lambda} \lambda 0 \lambda 0.

\smallskip
Rule 15-85
has  the mean-field approximation
$$
d_{t+1}=1-d_t.
$$
Again, the equation does not involve $\lambda$, for the same reason as before. It has a single fixed point $d=1/2$, in agreement with the experimental observations. It is worth noting, however, that although the steady-state density is independent of $\lambda$, the dynamics of this rule is still quite rich. Its space-time diagram, shown
in Figure~\ref{fig::stdZDCD}, is composed of vertical and horizontal stripes. The proportion  of horizontal vs. vertical stripes varies with $\lambda$, and we here have a symmetry-breaking phenomenon with a competition between the two patterns. For the time scales observed, we find that the density remains around $1/2$ but one can also predict that there is always a non-zero probability that the systems attains a cycle where the all-zero and the all-one configurations alternate. This phenomenon needs to be studied in more details.

We should also add that that there exists a rule in the CD class for which $d_\infty$ appears to be constant but different  from $1/2$, namely rule 3-17,
for which $d_\infty \approx 0.71$. Although the mean-field approximation equation for this rule has a fixed point $1/2$, which is clearly wrong, already the second-order local structure approximation yields very good agreement with 
our experiments (we only observed it numerically, as the
local structure equations are not explicitly solvable).


\subsection{Rules 10-245 (class LD) and  36-219 (class NLD)}
ECA rules 10 and 245 are the reflected-conjugated version of each other. The diploid 10-245 has the following transition probabilities

\vspace{4pt}
\tabP \lambda {1 - \lambda} \lambda {1 - \lambda} \lambda \lambda \lambda \lambda .

\smallskip
The mean-field equation is
$$
d_{t+1}= \left(2\,\lambda -1 \right) d_t^2- \left(2\,\lambda -1 \right) d_t+\lambda.
$$
It has a stable fixed point at
\begin{equation}\label{10-245MF}
d={\frac {\lambda-\sqrt {-{\lambda}^{2}+\lambda}}{2\,\lambda-1}},\end{equation}
which clearly is not a linear function of $\lambda$.
However, when we expand the above expression around 
$\lambda=1/2$, we obtain
$$
d=\frac{1}{4}+\frac{\lambda}{2}+ O \left(  \left( \lambda-1/2 \right) ^{3} \right). 
$$
The linear part of this expansion, $d=\frac{1}{4}+\frac{\lambda}{2}$, if plotted, appears to be almost indistinguishable for  experimental data.
This means that the mean-field approximation is quite good in the vicinity of $\lambda=1/2$, but deteriorates away from this point.

\begin{figure}
	\begin{tabular}{ c c }
		rule 10-245 (LD) & rule 36-219 (NLD)\\
		\includegraphics[width=0.47\linewidth]{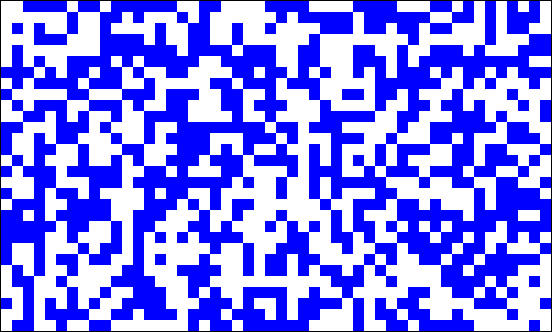}&
		\includegraphics[width=0.47\linewidth]{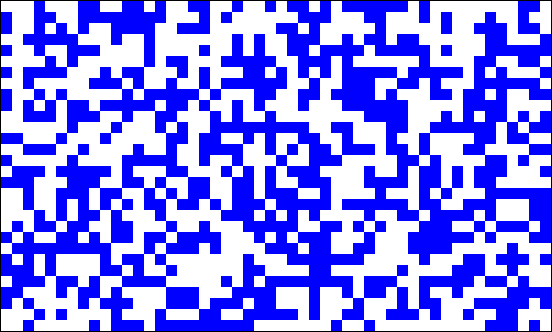}
	\end{tabular}
	
	\caption{Space-time diagrams for the rules 10-245 and 36-219 for $\lambda = 1/2$.}
	\label{fig::stdLDNLD}
\end{figure}

The local structure approximation of order 2 yields exactly the same expression as the mean-field, namely eq.~(\ref{10-245MF}), so there is no improvement.
However, when the order  is increased to $3$, it yields the expression
$$d=\frac{1}{4}+\frac{\lambda}{2}.$$
We could thus conclude that while the first- and second-order local structure approximation is not too bad, the third order becomes
practically indistinguishable from simulation data.  
This good approximation might be related to the disordered (or chaotic) evolution of the system as shown on Fig.~\ref{fig::stdLDNLD}-left. 
A possible origin of this agreement is that the local structure approximation constructs
a Bayesian extension of finite-block measures, and this extension
maximizes measure entropy. One would thus expect that very
the most ``disordered'' steady states are those which are well-approximated by
the local structure theory.

Let us now turn our attention to rule 36-219, also constructed from rules which are the reflected-conjugated versions of each other. 
Its transition probabilities are 

\vspace{4pt}
\tabP \lambda  \lambda {1 - \lambda}  \lambda \lambda {1 - \lambda} \lambda \lambda

\begin{figure}
 \begin{center}
  \includegraphics[width=9cm]{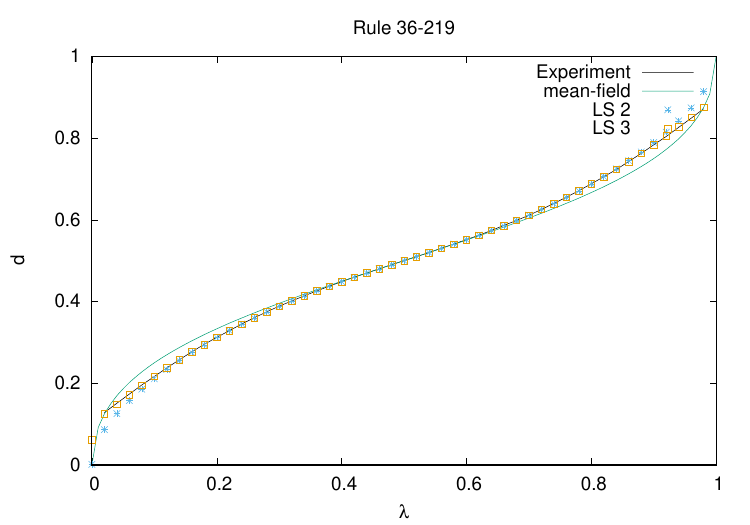}
 \end{center}
\caption{Graph of the steady-state density and its approximations
for rule 36-219.}\label{Fig-36-219}
\end{figure}

\smallskip
Its space-time diagram is shown on the right of Figure~\ref{fig::stdLDNLD}. 
Coincidentally, the mean-field approximation for rule 36-219
yields the same equation as for rule 10-245 discussed above.
The resulting graph is shown in Figure~\ref{Fig-36-219}.
Local structure approximation equations of the second order 
are not possible to solve explicitly, thus Figure~\ref{Fig-36-219} shows numerical results
only. One can see that the second-order approximation is better than the mean-field, but it departs from the experimental curve near $\lambda=0$ and $\lambda=1$. Nevertheless, the third order local structure approximation, also shown in Figure~\ref{Fig-36-219}, provides an excellent fit, practically indistinguishable from the experimental curve. The space-time diagrams are also very similar to those observed for rule 10-245 (see Fig.~\ref{fig::stdLDNLD}).

\subsection{Rules 60-102 and 46-116 (class PT2)}
\begin{figure}
	\begin{tabular}{ c c }
		rule 60-102 (LD) & rule 46-116 (NLD)\\
		\includegraphics[width=0.47\linewidth]{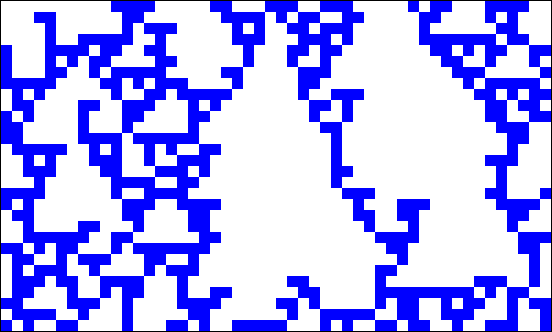}&
		\includegraphics[width=0.47\linewidth]{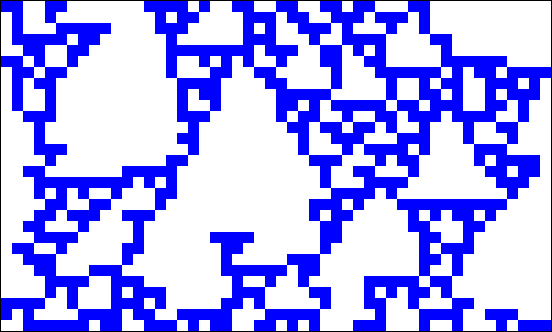}
	\end{tabular}

	\caption{Space-time diagrams for the rules 60-102 and 46-116 for $\lambda = 1/2$.}
	\label{fig::stdPT2}
\end{figure}

Let us now focus on rules where a second-order phase transitions appears.
For rules belonging to this class, PT2, we will discuss two
examples, namely the  endogamous rules 60-102 and 46-116. Note that elementary rules 60 and 102, as well as 46 and 116, are  reflected versions of each other.
\begin{figure}
 \begin{center}
  \includegraphics[width=9cm]{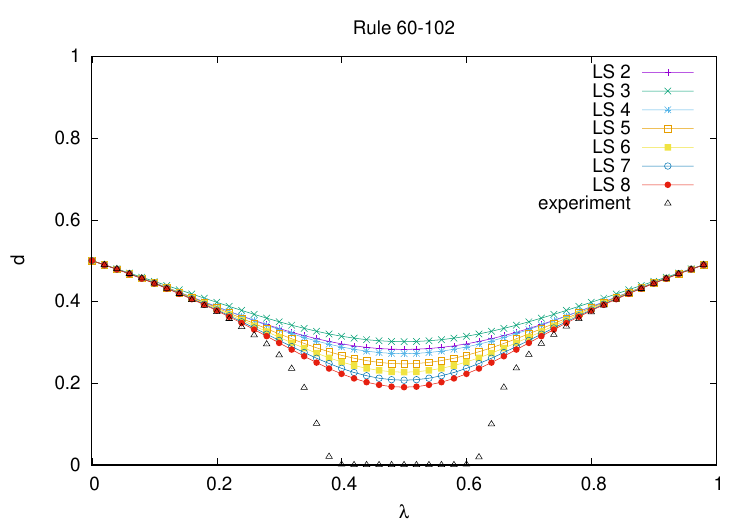} \includegraphics[width=9cm]{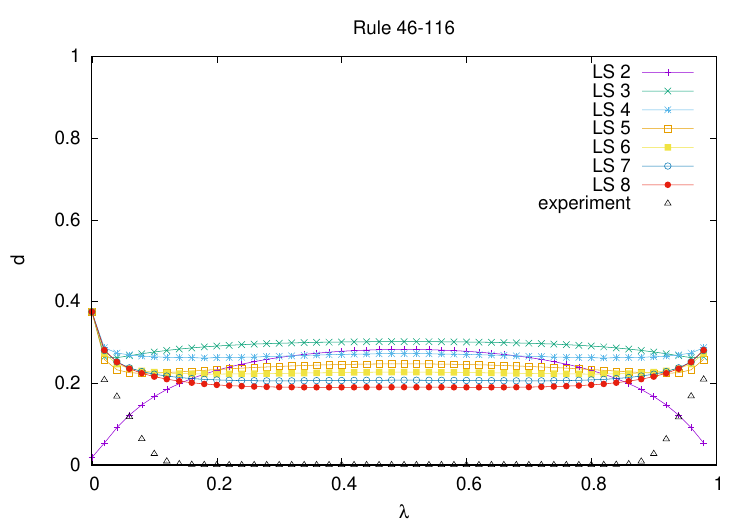}
 \end{center}
\caption{Graph of the steady-state density and its approximations
for rules 60-102 (top) and 46-116 (bottom).}\label{Fig-60-102}
\end{figure}
For rule 60-102, the transition probabilities are 

\vspace{4pt}
\tabP 0 \lambda 1 {1 - \lambda} {1 - \lambda} 1 \lambda 0 .

\smallskip

The mean-field equation is
$$d_{t+1}=-2d_t^2+2d_t,
$$
yielding the stable fixed point $d=1/2$. 
Local structure of order 2 yields
$$
d=\frac{
5 {\lambda}^{2}-5 \lambda+3-\sqrt {5\,{\lambda}^{4}-10\,{\lambda}^{3}+7\,{\lambda
}^{2}-2\,\lambda+1}}
{4\,{\lambda}^{2}-4\,\lambda+4}.
$$
The graph of this expression is shown in the top of Figure~\ref{Fig-60-102}
labeled as ``LS 2''. 
Local structure equations of higher order are not solvable analytically, but we show the numerical solutions obtained with orders up to 8 in Figure~\ref{Fig-60-102}. 

None of these
curves exhibit discontinuities of the derivative present in the experimental curve. They do exhibit the ``dip'' around the center,
which becomes deeper with the increasing order of the approximation,
but, surprisingly, even the eighth order is very far from the actual curve.
Cirillo \emph{et al.} actually considered orders up to 13 for the same rule (60-102), and still failed to observe discontinuity of the first derivative~\cite{CLS24}. 

\smallskip
For rule  46 -- 116,
the transition probabilities are

\vspace{4pt}
\tabP 0 {1 - \lambda} 1 {1 - \lambda} \lambda 1 \lambda 0 .

\smallskip
The mean-field equation for rule 46-116 is
exactly the same as for 60-102
with stable fixed point at $d=1/2$. The local structure
of order 2 gives 
$$
d=\frac{10\,{\lambda}^{2}-10\,\lambda-1 -\sqrt {4\,{\lambda}^{4}-8\,{\lambda}^{3}-4\,{\lambda
}^{2}+8\,\lambda+1}}{16\,{\lambda}^{2}-16\,\lambda-2},
$$
which, as we can see from the graph of the above expression shown in the bottom of Figure \ref{Fig-60-102} (with label ``LS 2''), is even worse than in the case of rule 60-102. Higher-order local structure approximations, numerically obtained, start resembling the U-shaped
graph only for order 5 and higher, but even then they remain much too far from the experimental curve. It appears that for this rule too, the local structure theory gives poor results.
Space-time diagrams of rules 60-102 and 46-116 are shown in Fig.~\ref{fig::stdPT2}.

The discontinuity of the derivatives for experimental curves observed in Figure~\ref{Fig-60-102} indicate the possible presence of a second-order phase transition, with the order parameter $d_\infty$ and the control parameter $\lambda$. Rules 60-102 and 46-116 have \emph{two} such transitions, one at $\lambda_c$ and the other at $1-\lambda_c$, due to the symmetry of graphs.

 In order to determine experimentally more precisely the critical value of $\lambda$, we performed another series of experiments, which consists in plotting the the density as a function of time in log-log scale near the critical value of $\lambda$. The theory of phase transitions predicts that when $\lambda$ is precisely at the critical value, the density converges toward 0 as a power law, with an exponent which is ``universal'' in the sense that it does not depend on the local rule but only on the general features of the model. More precisely, when the phase transition belongs to universality class of {\em directed percolation}, at the critical value of $\lambda$, one should observe $ d_t\sim t^{-\delta} $ with $\delta=0.159464(6)$ determined experimentally with advanced statistical physics techniques (see e.g., \cite{Hinrichsen01112000}).

\begin{figure}
	\begin{center}
		\includegraphics[width=0.48\linewidth]{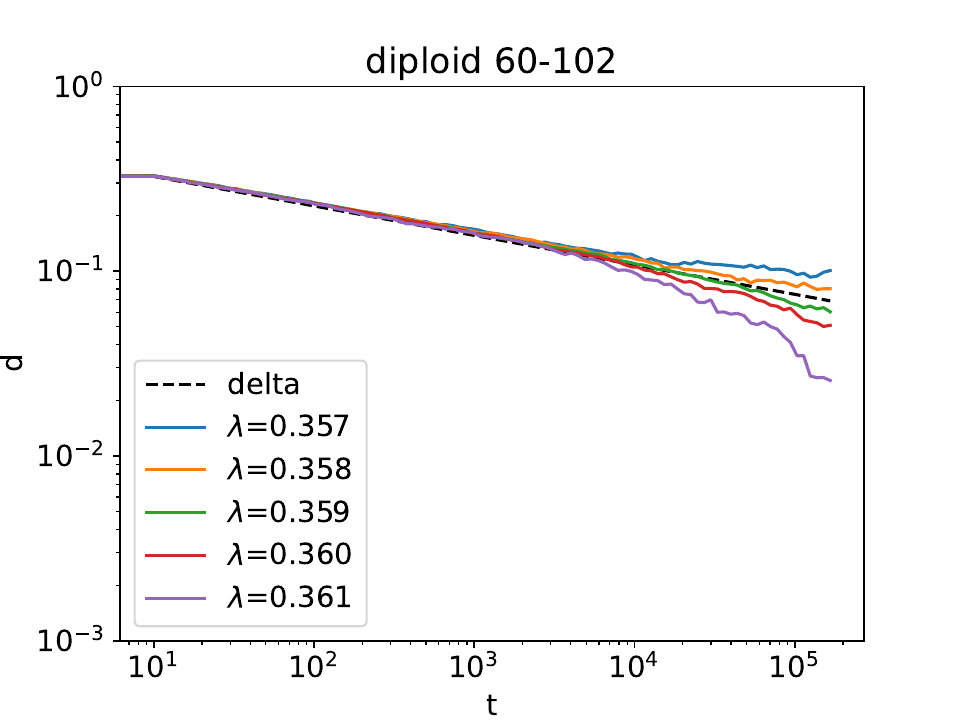}
		\includegraphics[width=0.48\linewidth]{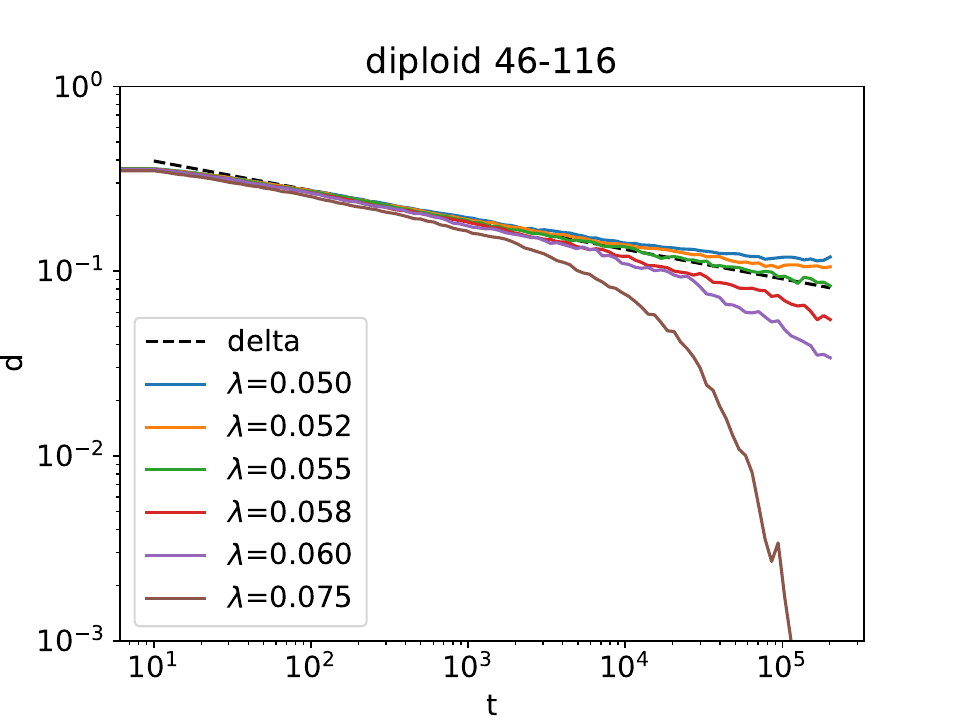}
	\end{center}
	\caption{Evolution of the density  as function of time in log-log scale. The dashed line has a slope $ \delta = -0.159464$, which corresponds to the critical exponent for directed percolation.}
	\label{fig::directedPercolation}
\end{figure}

 The resulting graphs for both rules 60-102 and 46-116 are shown in Fig.~\ref{fig::directedPercolation}. 
 These graphs are in good  agreement with the aforementioned prediction. We can observe that a change of curvature occurs near the critical value of $ \lambda$, with an almost straight-line behaviour in the log-log scale (the slope of the straight line shown in Fig.~\ref{fig::directedPercolation} as the dashed line is $-\delta$).
We found that for rule 60-102
$ \lambda_c \approx 0.359 $
and for rule 46-116
$ \lambda_c \approx 0.055.$ 
Recall that for symmetry reasons, for both rules, the absorbing phase, that is, the existence of an all-zero steady-state, lies in the intervals $ [\lambda_c,1-\lambda_c]$ and the active phases lie in the two other intervals.


\subsection{Rule 128-254 (class PT1)}

\begin{figure}
	\begin{tabular}{ c c c }
	\includegraphics[width=0.30\linewidth]{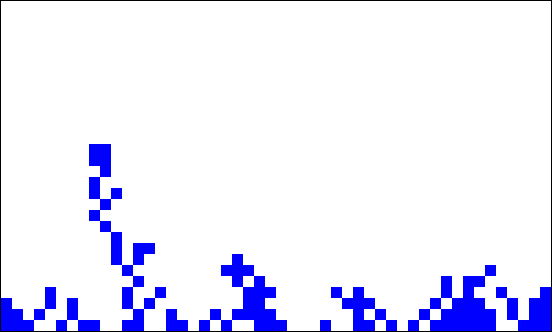}&
	\includegraphics[width=0.30\linewidth]{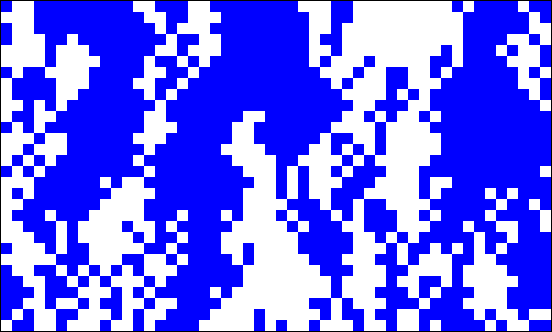}&
	\includegraphics[width=0.30\linewidth]{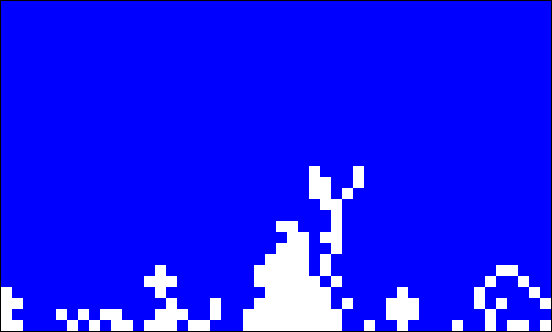}\\
\end{tabular}	
	\caption{Space-time diagram for rule 128-254: (left)  $\lambda = 1/4$, (middle)  $\lambda = 1/2$, (right)  $\lambda = 3/4$.}
	\label{fig::stdPT1}
\end{figure}
In the diploid rule 128-254, elementary rules 128 and 254 are the conjugated versions of each other; both are invariant by reflection.
The transition probabilities are
\vspace{4pt}

\vspace{4pt}
\tabP 0 \lambda \lambda \lambda \lambda \lambda \lambda 1 .

\begin{figure}
 \begin{center}
  (a)\includegraphics[width=8.7cm]{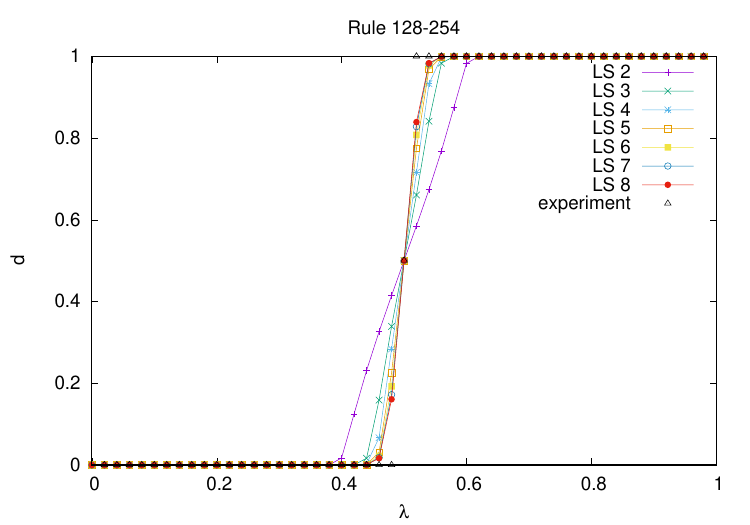}\\
 (b)\includegraphics[width=8.7cm]{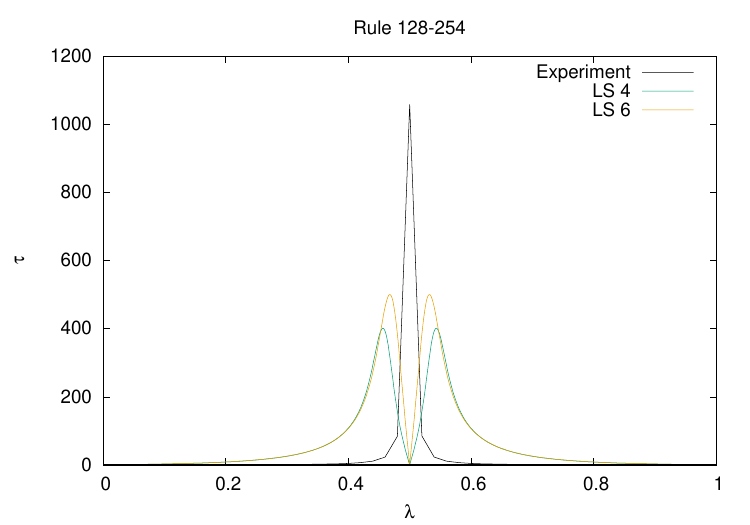}
 \end{center}
\caption{(a) Graph of the steady-state density and its approximations
for rule 128-254.
(b) Graph of the convergence time for the same rule.
}\label{Fig-128-254}
\end{figure}

\smallskip
The mean-field approximation equation for rule 128-254 is
$$
d_{t+1}=d_t^3 - 3 \lambda d_t^2+  3 \lambda d_t^3.
$$
It has three fixed points, $d \in \{0, 3\lambda-1, 1\}$, with respective intervals of stability $(0,1/3)$, $[1/3,2/3)$ and
$[2,3, 1)$. This means that the mean-field steady state 
is given by
\begin{equation}\label{slopedstep}
d=\begin{cases}
0 & \lambda \in (0, 1/3) \\
3\lambda -1 & \lambda \in [1/3, 2/3)\\
1 & \lambda \in [2/3, 1).
\end{cases}
\end{equation}
Local structure approximation of order 2 yields exactly the same
steady state, its graph is shown in Figure \ref{Fig-128-254}a.
This function is continuous, thus it does not accurately
represent the behaviour of $d_\infty$ vs. $\lambda$ observed experimentally, which is a step function with discontinuity at $\lambda=1/2$. However, when the order of the local structure approximation increases, the resulting graph seems to be
also piecewise-linear 
similarly to eq.~(\ref{slopedstep}), but with the middle part becoming steeper and steeper. One can conjecture that in the limit of infinite order,
$d$ will become
\begin{equation*}
d=\begin{cases}
0 & x\in (0, 1/2) \\
1 & x\in [1/2, 1).
\end{cases}
\end{equation*}

The discontinuity at $\lambda=1/2$ for rule 128-254 exhibits features
of a first-order phase transition (see Fig.~\ref{fig::stdPT1}). 
In particular, we can observe a decrease of the speed of convergence around $\lambda=1/2$.
 In order to illustrate this phenomenon, let us define the convergence time $\tau$ 
as
$$
\tau = \sum_{t=0}^\infty |d_t-d_{\infty}|.
$$
If $d_t$ converges to $d_{\infty}$ exponentially, this sum will be finite. If, on the other hand, $d_t$ converges to $d_{\infty}$ 
as a power law, $\tau$ will become infinite. In numerical simulations, instead of $d_{\infty}$ we take  $d_{T}$ with large $T$.
Figure~\ref{Fig-128-254}b shows the graph of 
the convergence time vs. $\lambda$. We can see that the experimental curve has a pronounced peak at $\lambda=1/2$.
This peak is finite because we use both finite lattice size and finite time $T$. Numerical evidence suggests that in the limit of $L\to \infty$
and $T \to \infty$ the height of this peak will go to infinity and its width will go to zero.
Figure~\ref{Fig-128-254}b also shows examples of $\tau$ vs. $\lambda$ curves obtained using the local structure approximation. The local structure graphs exhibit \emph{two} peaks, symmetrically located 
on both sides of $\lambda=1/2$. When the order of approximation 
increases, the height of these peaks increases and they get closer and closer to  $\lambda=1/2$. In the limit of infinite order these two peaks will merge.
 In this sense we can say that
the local structure approximation predicts the existence of the first-order phase transition, although in a somewhat ``distorted'' fashion.

\section{Solvable cases}\label{solvable}

Some endogamous diploid rules turn out to be special cases of 
a known PCA rule for which exact results exist. 
This PCA rule is defined below in Sec.~\ref{modeldef}, and 
in Sec.~\ref{modelexample1} and \ref{modelexample2}
we present two examples of rules that are special cases: one belonging to class ZD and one to class NLD.

\subsection{Description of the general PCA model}\label{modeldef}
Consider the probabilistic rule proposed as a simple model for diffusion of innovations, spread of rumours, or a similar process involving transport
of information between neighbours~\cite{hfpaper7}. We consider an infinite one-dimensional lattice where each cell is occupied by
an individual who has already adopted the innovation (1) or who has not adopted it yet (0). 
Individuals who adopted the innovation remain in state 1 forever. Individuals in state 0 change can their states to 1
(adopt the innovation) with probabilities depending on the state of nearest neighbours. All changes of states
take place simultaneously. This process can be formally described as a binary probabilistic CA with the following
transition probabilities,

\begin{equation}
	\label{adpodef}
\tabP 0 \alpha 1 1 \beta \gamma 1 1 ,
\end{equation}

\smallskip
where $\alpha, \beta, \gamma \in [0,1]$ are parameters.
In \cite{hfpaper54}, the author used the cluster expansion method to find the exact expression 
for the probability of occurrence of zero for an arbitrary cell  after $n$ iterations of this rule,
providing that one starts with Bernoulli initial distribution
with $d_0=P_0(1)=p$, $P_0(0)=1-p$, $p \in[0,1]$.
In the book \cite{hfbook2023}, more details of the derivation are provided and the  expression for the density is 
given as 
\begin{equation}\label{finalPdiffinn}		
 d_n = \begin{cases} 
\displaystyle  1-\theta^n (1+p)(1-p)^2
 - p^2(1-p) \tilde{a}^n  -\displaystyle \frac{K  (\tilde{a}^n-\theta^{n})}
 {\tilde{a}-\theta}    & \text{if }	\tilde{a} \neq \theta,\\
 1-(1-p) \tilde{a}^n - K \tilde{a}^{n-1}n & \text{if }	\tilde{a} = \theta,
 \end{cases}
\end{equation}
where
\begin{equation}\label{defconst}
\begin{aligned}
K&=b{ p ^2}(1-{ p })^2 +c { p ^2}(1-{ p })^3,\\
\theta&={a}+b(1-{ p })+c(1-{ p })^2,\\
 a&=(1-\alpha)(1-\beta), \\
 \tilde{a}&=1-\gamma,\\
  b&=\alpha-2\alpha\beta+\beta,\\
  c&=\alpha\beta.
\end{aligned}
\end{equation}

If $\alpha, \beta, \gamma$ are restricted to integer values in $\{0,1\}$, the rule becomes deterministic, and for 8 different combinations of parameters $\alpha, \beta, \gamma$ we find 8 different deterministic ECA. These rules, however, are not ''minimal'' in the sense of Table~\ref{minimalrules}. 
Since we would like to obtain
an endogamous diploid rule $W_f-W_g$
in which $f$ is a minimal representative rule, it is more convenient to consider
a Boolean conjugate of the rule given by eq.~(\ref{adpodef}).

If we replace 0's by 1's and 1's by 0's in the definition of transition probabilities, we find 

\begin{equation}
\begin{aligned} 
 w(0|111)&=0,\, w(0|110)=\alpha,\,w(0|101)=1,\,w(0|100)=1,\\
 w(0|011)&=\beta,\,w(0|010)=\gamma,\,w(0|001)=1,\,w(0|000)=1, 
\end{aligned}
\end{equation}
and, after sorting it back and using $w(1|xyz)=1-w(0|xyz$), 

\begin{equation}
\label{conj-innovation}
\tabP 0 0 {1-\gamma} {1-\beta} 0 0 {1-\beta} 1.
\end{equation}

\smallskip
Since this is the Boolean-conjugate rule to the rule defined in eq.~(\ref{adpodef}), we can obtain the expression for $ d_n$  by replacing, in eq.~(\ref{finalPdiffinn}), 0 by 1 and $p$ by $1-p$, 
\begin{equation}\label{finalPdiffinn-conj}		
d_n = \begin{cases} 
\displaystyle  \theta^n (2-p)p^2 
 + p(1-p)^2 \tilde{a}^n  +\frac{K  (\tilde{a}^n-\theta^{n})}
 {\tilde{a}-\theta}    & \text{if }	\tilde{a} \neq \theta,\\
 p \tilde{a}^n + K \tilde{a}^{n-1}n & \text{if }	\tilde{a} = \theta.
 \end{cases}
\end{equation}
All the other symbols remain defined as in eq.~(\ref{defconst}).

\subsection{Rule 136-192 (class ZD)} \label{modelexample1}

\begin{figure}
	\begin{tabular}{ c  c }
		rule 136-192 (ZD) & rule 140-196 (NLD)\\
		\includegraphics[width=0.47\linewidth]{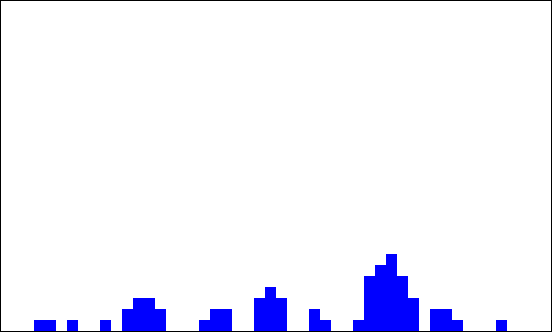}&
		\includegraphics[width=0.47\linewidth]{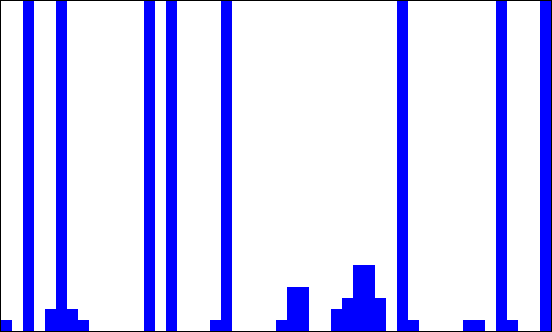}\\	
	\end{tabular}
	\caption{Space-time diagram for rule 136-192 (left) and 140-196 (right) for $\lambda=1/2$.}
\label{fig::stdZD}
\end{figure}

Consider now the two rules $136=(10001000)_2$
and $192=(11000000)_2$. 

Taking $\alpha=1-\lambda$, $\beta=\lambda$, $\gamma=1$, eq.~(\ref{conj-innovation}) becomes

\vspace{4pt}
\tabP 0 0 0 {1-\lambda} 0 0 \lambda 1 .

\smallskip
For $\lambda=0$, this defines rule 136, and for $\lambda=1$, rule 192, which means that these two rules are extremum points of the model defined in eq.~(\ref{conj-innovation}).
For general $\lambda \in [0,1]$, therefore, this is the diploid rule 136-192.
For our values of parameters, we find $\tilde{a}=0$ and $\theta=-\lambda^2/4 +\lambda/4+1/2$, thus $\tilde{a} \neq \theta$ and only the first line of 
eq.~(\ref{finalPdiffinn-conj}) applies. For $p=1/2$, 
eq.~(\ref{finalPdiffinn-conj}) simplifies to
$$
d_n=
\frac{3}{8}\theta^n+K \theta^{n-1}.
$$
Since $0<\theta<1$ for all $\lambda \in [0,1]$, we obtain
$d_\infty=0$,
in agreement with the numerical observations (see Fig.~\ref{fig::stdZD}).

We add that the mean-field equation for this rule is
$
d_{t+1}=d_t^2.
$
It has two fixed points 0 and 1, but only 0 is stable. 
This means that the mean-field approximation
yields the exact steady-state density, similarly as for the  rule 
2-16 which we considered earlier, also belonging to the ZD class. 

\subsection{Rule 140-196 (class NLD)} \label{modelexample2}
As a second case, consider rules $140=(10001100)_2$ and
$196=(11000100)_2$.
Take $\alpha=1-\lambda$, $\beta=\lambda$, $\gamma=0$.
Eq.~(\ref{conj-innovation}) becomes
\vspace{4pt}

\tabP 0 0 1 {1-\lambda} 0 0 \lambda 1.

\begin{figure}
 \begin{center}
  \includegraphics[width=9cm]{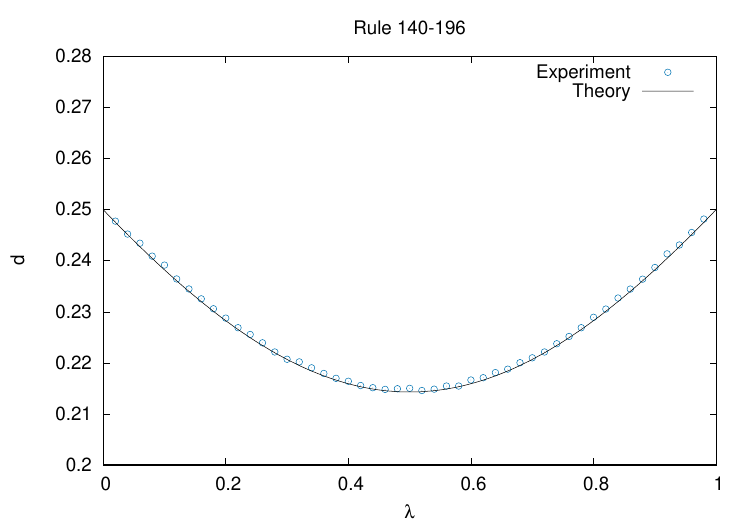}
 \end{center}
\caption{Graph of the steady-state density and its theoretical value for rule 140-196.}\label{Fig-140-196}
\end{figure}

\smallskip
For $\lambda=0$, this defines rule 140, and for $\lambda=1$, rule 196, thus again for $\lambda \in [0,1]$  we have the diploid rule 140-196.  Now we find $\tilde{a}=1$, and all the other parameters are the same as in the previous case.  
For $p=1/2$, 
eq.~(\ref{finalPdiffinn-conj}) simplifies to
$$
d_n=
\frac{3}{8}\theta^n+\frac{1}{8} +K \frac{1-\theta^{n}}{1-\theta}.
$$
Since again, $\theta^n \to 0$ as $ n \to \infty$, we find
$$
d_\infty=
\frac{1}{8} +\frac{K}{1-\theta}.
$$
After computing $K$ and $\theta$,
$$
K=-\frac{3}{32} \lambda (1-\lambda)+\frac{1}{16},
\quad
\theta = \frac{1}{4} \lambda (1-\lambda)+\frac{1}{2},
$$
we obtain, after simplification,
\begin{equation}\label{sol140-196}
d_\infty=
\frac{1}{2}
\frac{\lambda^2-\lambda+1}{\lambda^2-\lambda+2}.
\end{equation}
This agrees with observed numerical values, as illustrated in Figure~\ref{Fig-140-196}. The small deviations of the experimental points from the theoretical curve are due to the fact that
the numerical values were obtained for a finite lattice size and a finite number of iterations, yet eq.~(\ref{sol140-196}) was derived assuming an infinite lattice size and taking $n\to \infty$ limit.

The mean-field map for this rule has two fixed points 0 and 1, thus it does not yield the correct prediction. Higher-order 
local structure approximations are not solvable, yet
numerically the third-order one produces a steady state
which is very close to the exact solution. The convergence is even faster than in the case of rule 36-219 which we considered earlier (also belonging to the NLD class).

\section{Conclusions}
We examined the dependence of the asymptotic density
on the parameter $\lambda$  in diploid cellular automata, focusing on the subset of endogamous rules. We proposed to divide the density vs. $\lambda$ curves
into six distinct classes and studied these classes with some detailed examples. 
These six classes are a convenient way to 
investigate how $d_\infty$ depends on $\lambda$ and, to date, we do not have an algorithm which would decide to which class a given diploid rule belongs.
However, for each class, the selected examples allowed us to estimate how well the local structure predicts the asymptotic density as a function of $\lambda$.

The presented evidence suggests that for rules for which the steady-state density is a linear function of $\lambda$ (our classes ZD, CD and LD), the finite-order local structure approximation correctly reproduces the graph
of $d_\infty$ vs. $\lambda$.
 For rules with  differentiable but not linear graphs (class NLD),
the local structure either becomes exact at some finite order or converges rapidly to the exact solution as the order of the approximation increases.  

For rules with the first-order phase transition (class PT1),
the local structure does not produce a discontinuous steady state
for finite orders,
yet with the increasing order the
graphs seem to converge toward the sharp step function, thus correctly
reproducing the first-order phase transition phenomenon.

The good predictions of the local structure approximation
have one exception: for rules with second-order phase transitions (class PT2), even higher-order local structure equations do not appear to exhibit any bifurcation. Moreover, the predicted $d_\infty$ vs. $\lambda$ graphs only vaguely resemble the shape of the actual graph. This was already reported 
by Cirillo et al. for rule 60-102~\cite{CMP26}, but we confirmed that this is also the case  for rule 46-116. The phase transitions observed for both rules 60-102 and 46-116 belong to the directed percolation class.
It is an open question to understand why the local structure theory here fails while it gave reasonably accurate results in other cases~\cite{hfpaper49}.

 This is in a striking contrast with other stochastic rules investigated by the authors earlier, namely $\alpha$-asynchronous CA \cite{hfpaper49}.
 In $\alpha$-asynchronous rules with second-order phase transitions belonging to the directed-percolation (DP) universality class, even low order local structure approximation exhibits transcritical bifurcations. This suggests that the double phase transitions in endogamous diploid CA are very different than many other PCA rules belonging to the directed percolation universality class. They clearly deserve further study in order to explain the origin of this phenomenon. 

We will finish remarking that numerous open problems remain. For example,
for rule  10-245, the expression for the asymptotic density,
$d=\frac{1}{4}+\frac{\lambda}{2},$
appears to be exact. Could this be formally demonstrated?
Similarly, for rule 39-219, the third order seems to produce the graph of $d_\infty$ vs. $\lambda$ which is nearly indistinguishable from the experimental
graph. Does it mean that this approximation is actually exact in this case?

\subsection*{Acknowledgments}
H. F.  acknowledges  the support of the Natural Sciences and Engineering Research Council of Canada (NSERC) as well as the support provided by the Simon Fraser University Research Computing Group (Fir) and the Digital Research Alliance of Canada (alliancecan.ca).

\bibliographystyle{alpha}
\bibliography{diploidca.bib}

\end{document}